\documentclass[%
superscriptaddress,
twocolumn,
nofootinbib,
amsmath,amssymb,
aps,
prd,
]{revtex4-1}

\usepackage[dvipsnames]{xcolor}
\usepackage[unicode]{hyperref}
\hypersetup{colorlinks=true, citecolor=MidnightBlue,
            linkcolor=MidnightBlue, urlcolor=MidnightBlue, linktocpage=true}
\usepackage{tabularx}
\usepackage{graphicx}
\usepackage{amssymb,amsmath,bm,tensor,braket}
\usepackage[varg]{txfonts}
\usepackage{enumerate}
\usepackage{mathtools}
\usepackage{tensor}
\usepackage[capitalize]{cleveref}
\usepackage[utf8]{inputenc}
\usepackage[normalem]{ulem}
\usepackage{dcolumn}
\usepackage{soul}
\usepackage{subfigure}
\usepackage[normalem]{ulem}
\usepackage{booktabs} 
\usepackage{comment}
\usepackage{siunitx}
\usepackage{mathrsfs}
\usepackage[T1]{fontenc}
\newcommand{\dd}{\mathrm{d}}
\newcommand{\ii}{\mathrm{i}}
\newcommand{\ee}{\mathrm{e}}
\newcommand{\p}{^{\prime}}

\newcommand{\modi}[1]{{\color{black}#1}}

\begin{document}
	
\preprint{APS/123-QED}

\title{Black hole quasinormal mode resonances}
	
\author{Yiqiu Yang}\email{yueyang1812@gmail.com}
\affiliation{Department of Physics, School of Physics, Peking University, Beijing 100871, China}%
\affiliation{Kavli Institute for Astronomy and Astrophysics, Peking University, Beijing 100871, China}

\author{Emanuele Berti}\email{berti@jhu.edu}
\affiliation{William H. Miller III Department of Physics and Astronomy, Johns Hopkins University, Baltimore, Maryland 21218, USA}

\author{Nicola Franchini}
\email{nicola.franchini@tecnico.ulisboa.pt}
\affiliation{CENTRA, Departamento de F\'{\i}sica, Instituto Superior T\'ecnico -- IST, Universidade de Lisboa -- UL, Avenida Rovisco Pais 1, 1049-001 Lisboa, Portugal}

\date{\today}
\begin{abstract}
  Black hole quasinormal mode frequencies can be very close to each other (``avoided crossings'') or even completely degenerate (``exceptional points'') when the system is characterized by more than one parameter.
  We investigate this resonant behavior and demonstrate that near exceptional points, the two modes are just different covers of the same complex function on a Riemann surface.
  We also study the characteristic time domain signal due to the resonance in the frequency domain, illustrating the analogy between black hole signals at resonance and harmonic oscillators driven by a resonant external force. 
  \modi{We consider a specific toy model displaying a resonance between the fundamental mode and the first overtone, and we find that taking into account the linear growth in time due to the resonance is necessary to accurately recover the quasinormal mode frequencies.}
\end{abstract}
	
\maketitle
	
\allowdisplaybreaks
	
\noindent \textbf{\em Introduction.}
In black hole (BH) perturbation theory~\cite{Chandrasekhar:1985kt}, the oscillations of a BH spacetime are characterized by a superposition of damped exponentials with complex frequencies $\omega_n$, the so-called quasinormal modes (QNMs)~\cite{Kokkotas:1999bd,Nollert:1999ji,Berti:2009kk}. A study of these damped oscillating modes leads to the conclusion that BH spacetimes are (modally) stable at first order in the perturbations~\cite{Regge:1957td,Zerilli:1970wzz,Vishveshwara:1970zz,Cunningham:1978zfa,Teukolsky:1973ha,Press:1973zz}. The QNM spectrum contains information on the geometric structure~\cite{Davis:1971gg,1972ApJ...172L..95G,Cardoso:2008bp,Cardoso:2016rao} and perhaps even the quantum nature~\cite{Hod:1998vk,Motl:2003cd,Andersson:2003fh,Oshita:2023cjz} of a BH spacetime. The goal of the BH spectroscopy program is to infer the properties of the BH remnant resulting from a merger~\cite{Dreyer:2003bv,Berti:2005ys,Baibhav:2023clw}, environmental effects\modi{~\cite{Leung:1997was,Barausse:2014tra,Cardoso:2024mrw,Yang:2024vor,Ianniccari:2024ysv,Laeuger:2025zgb}}, and possibly even modifications of general relativity~\cite{Berti:2015itd,Maselli:2023khq,Volkel:2022khh,McManus:2019ulj,Cardoso:2019mqo,Cano:2024ezp,Cano:2024jkd} through the observation of ringdown events in gravitational-wave detectors~\cite{LIGOScientific:2016lio,LIGOScientific:2021sio}.

The study of the QNM spectrum of Kerr and Kerr-Newman BHs revealed the presence of avoided crossings (or eigenvalue repulsions), a phenomenon occurring quite generically for any set of angular multipole numbers $(\ell,m)$~\cite{Onozawa:1996ux,Cook:2014cta,Dias:2021yju,Dias:2022oqm,Motohashi:2024fwt,Lo:2025njp}. Further, for massive scalar perturbations of a Kerr BH, two QNMs can even become completely degenerate at so-called ``exceptional points'' (EPs) for specific values of the BH's angular momentum $a/M$ and of the scalar field mass $M \mu$, where $M$ is the BH mass~\cite{Cavalcante:2024swt,Cavalcante:2024kmy}.
Interestingly, the system exhibits hysteresis: if we follow adiabatically a QNM in parameter space, the result will depend on the path. More specifically: if we let $(a/M,\,M \mu)$ change continuously along a closed path $\gamma$ that surrounds the EP in the parameter space, after one loop the fundamental mode will turn into the first overtone, and vice versa. In other words, if the QNMs depend on a vector of parameters $\boldsymbol{p}=\left\{p_{i}\right\}$, the observed hysteresis implies that the integral 
$
\oint_{\gamma} \nabla_{\boldsymbol{p}}\omega_{0}\cdot \dd\boldsymbol{p} =\omega_{1}-\omega_{0}\neq 0
$.
By Stokes' theorem, the only possibility is that the path $\gamma$ encloses a singularity.

In this Letter we address two questions: (i) what is the mathematical structure of QNMs around EPs? (ii) Does it have observational implications for time-domain waveforms?

We investigate the mathematical structure of EPs in the frequency domain and prove that EPs are characterized by a Riemann surface structure consistent with the intuitive understanding of Ref.~\cite{Cavalcante:2024kmy}. 
Moreover, we find that resonances between QNMs must lead to a linear growth in time in the waveform. This is analogous to the secular resonance phenomenon affecting harmonic oscillators in classical mechanics, when a periodic driving force at the frequency of the normal mode of the system triggers a resonance. 
We further show that if we insist on fitting the time domain waveform by an ordinary QNM superposition, the presence of a resonance leads to biased estimates of the fundamental QNM, while fits allowing for the theoretically predicted linear growth in time perform much better.
Throughout this work we use geometrical units ($G=c=1$) and set $M=1$.

\begin{figure*}
  \begin{minipage}{0.296\linewidth}
    \centering
    \includegraphics[width=0.975\textwidth]{"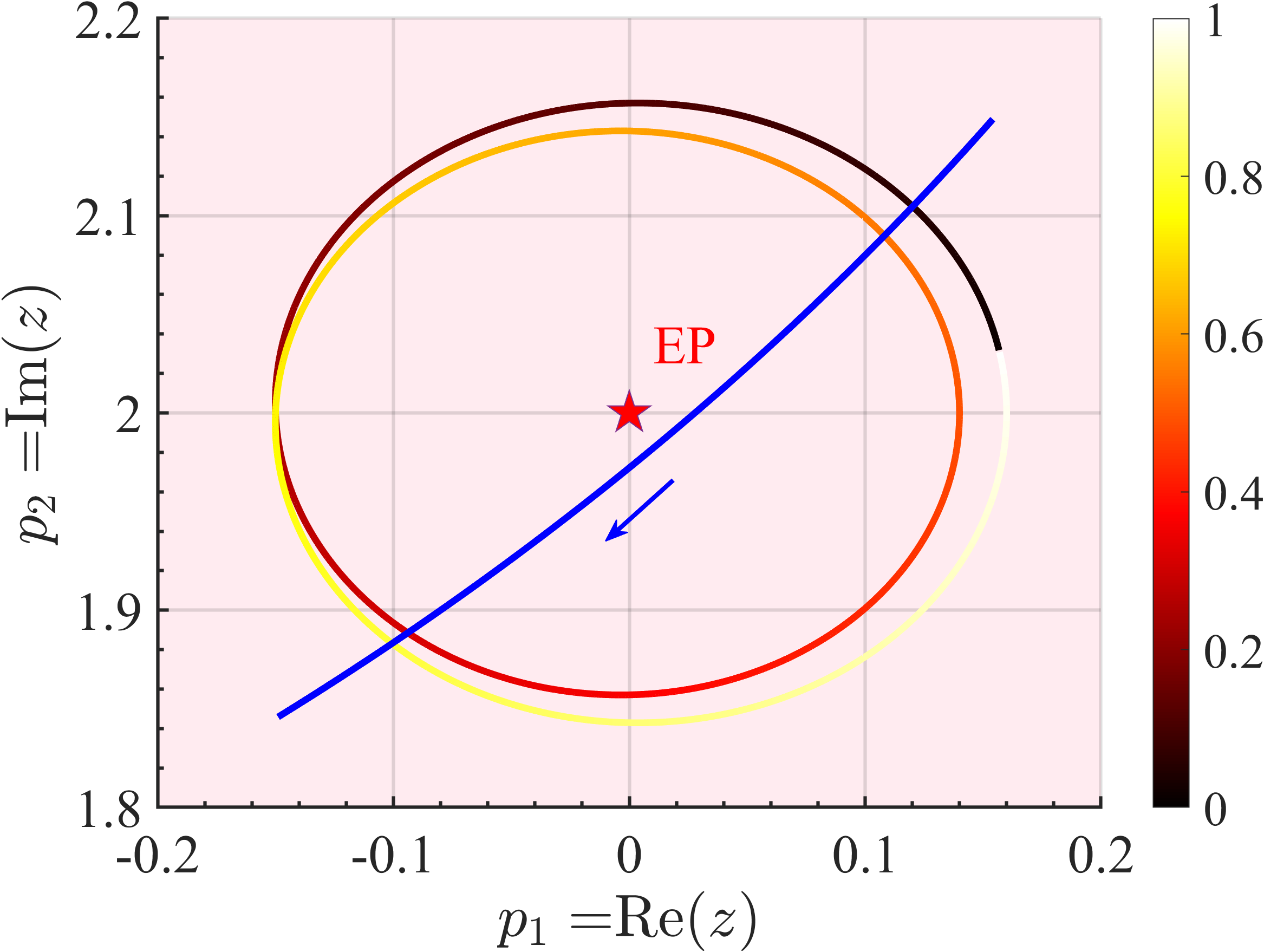"}
  \end{minipage}
  \begin{minipage}{0.212\linewidth}
    \centering
    \includegraphics[width=0.975\textwidth]{"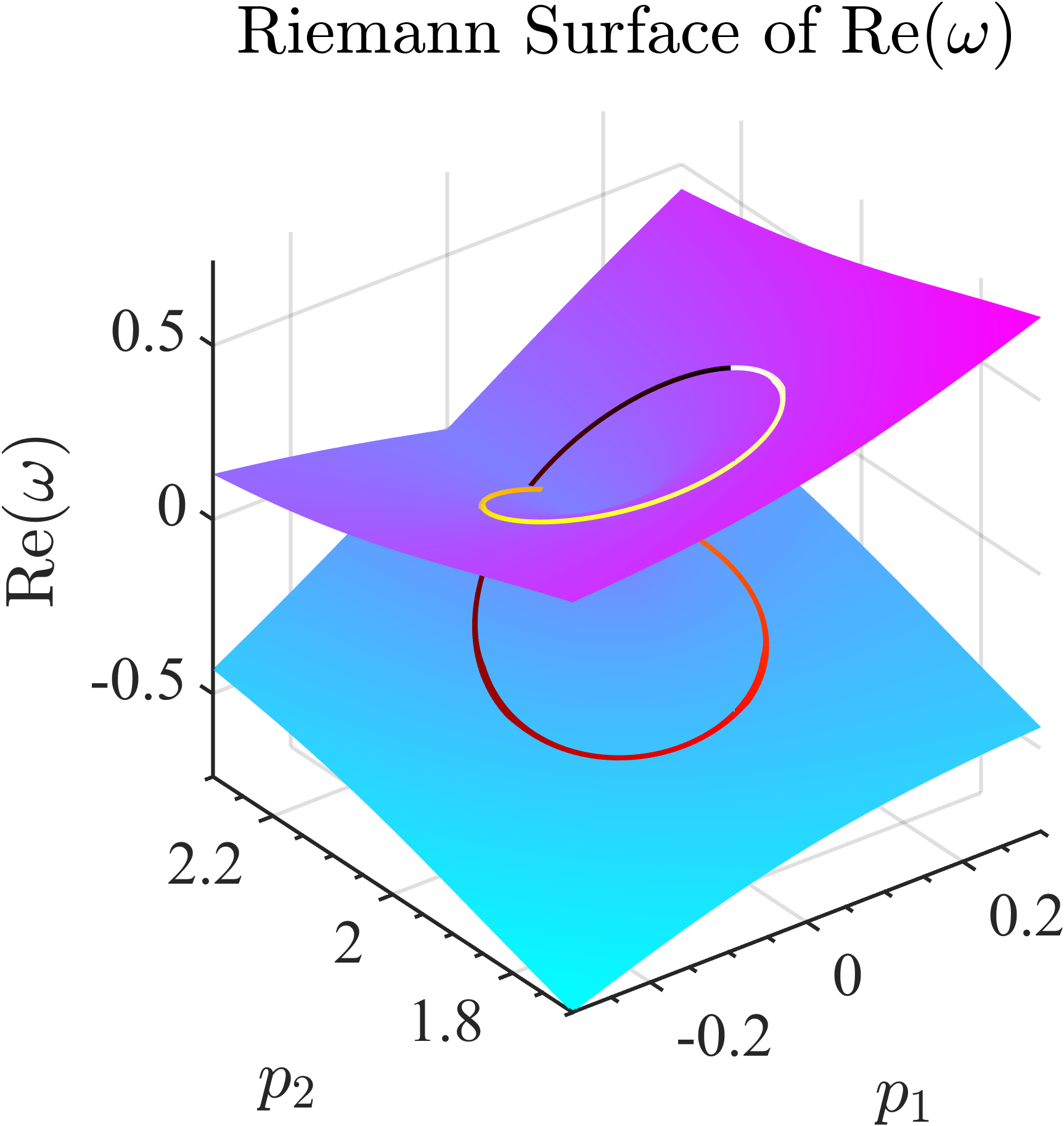"}
  \end{minipage}
  \begin{minipage}{0.232\linewidth}
    \centering
    \includegraphics[width=0.975\textwidth]{"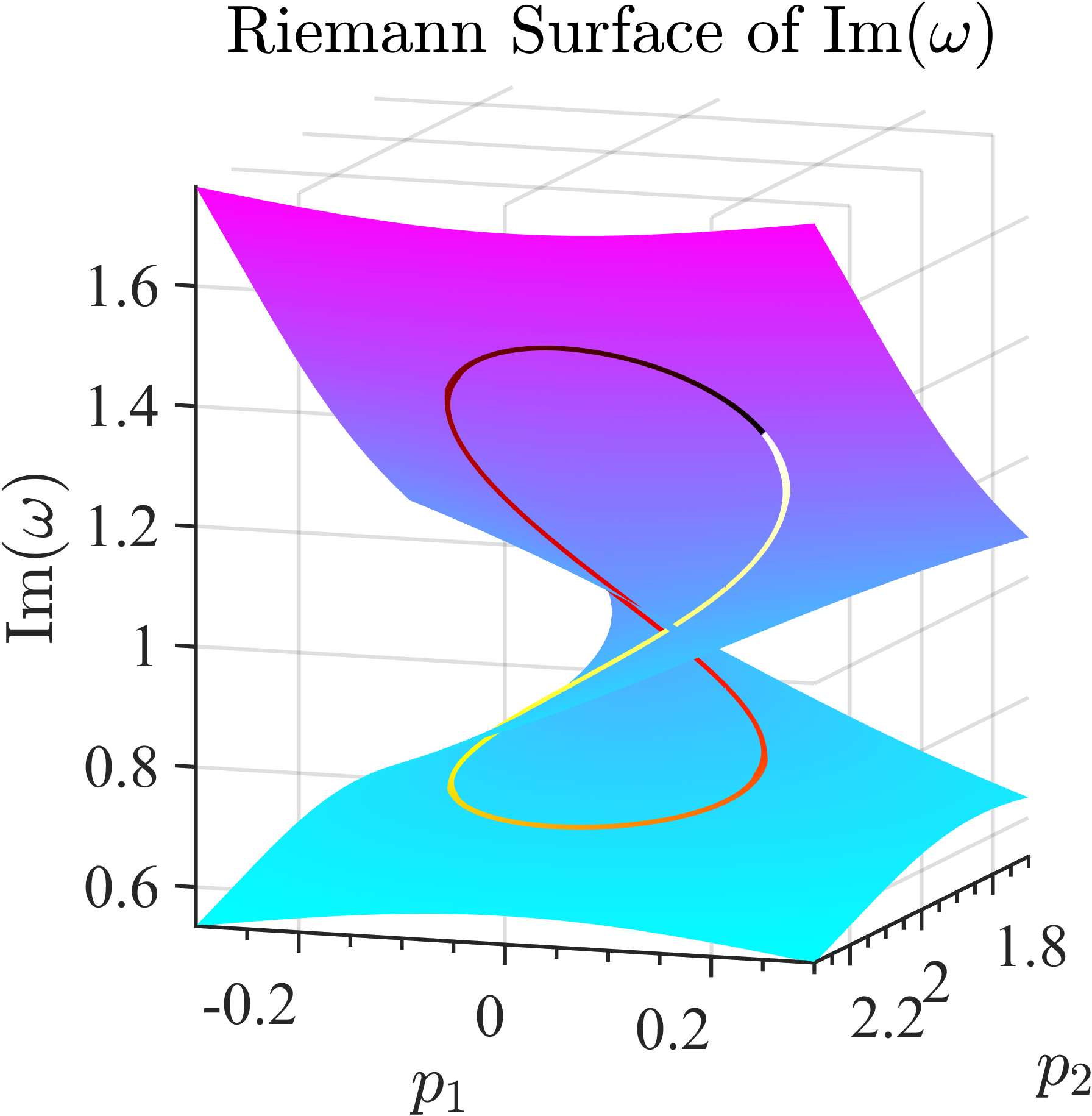"}
  \end{minipage}
  \begin{minipage}{0.222\linewidth}
    \centering
    \includegraphics[width=0.975\textwidth]{"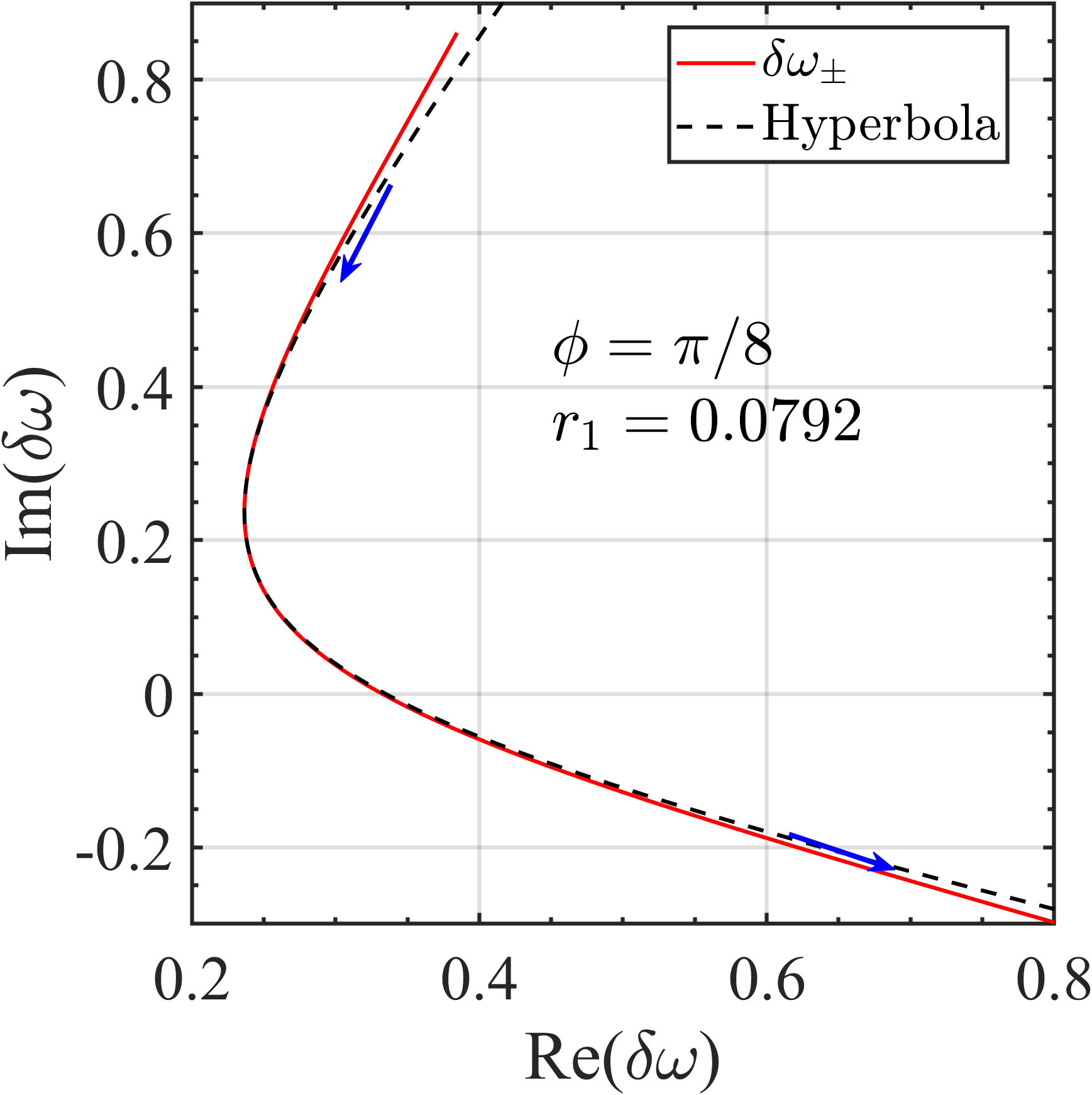"}
  \end{minipage}
  \caption{Schematic representation of Riemann surfaces and avoided crossings. Since the main features are universal, for simplicity we use the two-level system toy model described in the Supplemental Material~\cite{SM_Ref}. The system contains two free parameters denoted by $p_1$ and $p_2$, corresponding to the real part and imaginary part of a complex variable $z$. In the two central panels we show how the real and imaginary parts of the eigenvalues change as we complete a loop in the parameter space (leftmost panel): after one loop around the EP we do not recover the original eigenvalue. The blue line in the leftmost panel shows a characteristic path for eigenvalue repulsion, and the corresponding $\delta\omega_{\pm}=\omega_{+}-\omega_{-}$ is shown by the red line in the rightmost panel. The black dashed line is a hyperbola, that fits the trajectory of $\delta\omega_{\pm}$ very well~\cite{Motohashi:2024fwt}.}
  \label{fig:riemannsurface}
\end{figure*}

\noindent \textbf{\em Exceptional points.}
Perturbations of Kerr BHs are governed by a single master equation~\cite{Teukolsky:1973ha}. After separation of variables, the equation can be decomposed into
\modi{a set of angular and radial equations that depend on angular numbers $(\ell,\,m)$}.
The QNM frequencies
\modi{$\omega_{\ell mn}$} for a given effective potential are defined as ``free'' BH oscillations that correspond to (first-order) zeros in the amplitude of the incoming waves at infinity (see the Supplemental Material~\cite{SM_Ref} for details). 
In other words, we look for the zeros $\omega_{n}$ of an analytic complex function of $\omega$ in the frequency domain:  
\begin{equation}\label{qnmcondition}
  A_{\text{in}}(\omega_{n}  ,p_{i})=0\,,
\end{equation}
where \modi{we will neglect the dependence on $(\ell,\,m)$ for convenience from now on as the arguments are the same among spectrums with different $(\ell,\,m)$. In the equation above,} $p_{i}$ denotes the parameters that govern the perturbed BH: these would be e.g. $(a/M,\,M \mu)$ for massive scalar perturbations of a Kerr BH, or the spin $a/M$ and the coupling constant $\lambda$ of any additional terms in the Lagrangian for higher derivative modified gravity theories~\cite{Cano:2024ezp}.

For simplicity, we first illustrate general properties of EPs in the two-parameter case (later on we will comment on extensions to a higher-dimensional parameter space). 
As we mentioned earlier, there must be a singularity within any closed loop surrounding an EP. 
Here we argue that the singularity is the EP (see the Supplemental Material~\cite{SM_Ref} for a more detailed proof).
Using Eq.~(\ref{qnmcondition}), we can write down the gradient
\begin{equation}\label{migration}
  \nabla_{\boldsymbol{p}}\omega_{n}
  =-\dfrac{1}{A_{\text{in}}^{\prime}(\omega_{n},p_{i})}\nabla_{\boldsymbol{p}}A_{\text{in}}(\omega_{n},p_{i}) \,,
\end{equation}
where a prime denotes a partial derivative with respect to $\omega$. Consider two modes that become degenerate at the EP $\boldsymbol{p}=\boldsymbol{p_{\star}}$, i.e., $\omega_{n}(\boldsymbol{p_{\star}}) = \omega_{m}(\boldsymbol{p_{\star}}) = \omega_\star$. The coincidence of the two modes implies that $\omega_\star$ becomes a second-order zero of $A_{\text{in}}$ and a first-order zero of $A_{\text{in}}\p$, so the gradient becomes singular at this point.
Furthermore, by defining $\delta\omega_{nm}=\omega_{n}-\omega_{m}$, we can prove that around the EP
\begin{equation}\label{2foldfunction}
  \delta\omega_{nm}=\sqrt{\boldsymbol{A_{nm}}\cdot(\boldsymbol{p}-\boldsymbol{p_{\star}})}\,,
\end{equation}
where $\boldsymbol{A_{nm}}$ is a constant vector proportional to $\nabla_{\boldsymbol{p}}\omega_{n}(\omega_{\star},p_{i\star})$, which is generally nonzero. Therefore, around the EP, $ \delta\omega_{nm} $ behaves like the complex function $ z^{1/2} $ near the origin of the complex plane: a two-parameter EP gives rise to a two-sheeted Riemann surface, as shown for a simple toy model (described in the Supplemental Material~\cite{Heiss:2012dx} or other works~\cite{Heiss:2012dx,Ashida:2020dkc,Ding:2022juv}) in Fig.~\ref{fig:riemannsurface}. Whenever the parameters change along a closed trajectory $ \gamma(t) $ which encloses at most one EP, $ \delta\omega_{nm} \sim [\rho(t)\ee^{i\theta(t)}]^{1/2} $. After one loop (i.e., when the angle $\theta(t)$ changes by $ 2\pi $ and we get back to the starting point) $\delta\omega_{nm}$ picks up a minus sign, thus $\omega_{n}$ and $\omega_{m}$ get swapped, as first noticed numerically in Ref.~\cite{Cavalcante:2024swt}. Mathematically, the two complex functions $\omega_{n}(p_{i})$ and $\omega_{m}(p_{i})$ are just covers of the same complex function belonging to different Riemann sheets. By labeling overtones, we pick up a special choice of the branch cut, lying on the line where the modes' imaginary parts coincide (this is not the only way to define the branch cut, though it is physically plausible). 
	
If we have only one free parameter (e.g., $a/M$ for Kerr BHs), we cannot expect to find EPs, as the coincidence of two complex modes requires the fine-tuning of two free parameters. However, eigenvalue repulsion was observed for the Kerr case~\cite{Motohashi:2024fwt}. This can be understood with the help of Eq.~\eqref{2foldfunction}. Imagine a system where all parameters except for $p_1$ are fixed near the EP, as in the blue line in the first panel of  Fig.~\ref{fig:riemannsurface}. Thus, Eq.~\eqref{2foldfunction} can be rewritten as $\delta\omega_{nm}=\sqrt{c_{1}+c_{2}p_{1}}$, where $c_{1}$, $c_{2}$ are two complex constants. To see the form of the trajectory, we can further redefine the parameter $p_{1}$ to be $p_{1}\p=p_{1}+\text{Re}(c_{1}/c_{2})$, rewriting the equation into $\delta\omega_{nm}=\ee^{\ii\phi}$$\sqrt{r_{1}+\ii r_{2} p_{1}\p}$, where $r_{1}$, $r_{2}$ are real numbers and $\phi$ is some phase. We can define $\delta\omega_{nm}=\ee^{\ii\phi}(x+\ii y)$ and take the square of both sides to get $x^2-y^2=r_{1}$, $2xy=r_{2}p_{1}\p$. 
The fact that QNMs exhibiting eigenvalue repulsion describe hyperbolas in the complex plane close to the avoided crossing was indeed observed in Ref.~\cite{Motohashi:2024fwt}.
A specific example using a simple toy model is shown in Fig.~\ref{fig:riemannsurface}. 
	
When the vector $\boldsymbol{p}$ has $D_{p}>2$ components, in general we expect the resonance between two modes to live on a hypersurface of dimension $D_{p}-2$. The intersections of these hypersurfaces will create higher-order EPs, whose theoretical and physical implications have been studied extensively in other areas~\cite{2016PhRvX...6b1007D,Hodaei:2017mtx,Miri:2019fuk,Guo:2022rza,Kim:2023xdn,PhysRevLett.123.213901}. In general, on a ($D_{p}-2$)-dimensional exceptional hypersurface there will be $D_{p}$ coincident modes, with the splitting between these coincident modes scaling like $\left[\boldsymbol{A_{nm}}\cdot(\boldsymbol{p-p_{\star}})\right]^{1/D_p}$~\cite{Bergholtz:2019deh,Ding:2022juv,Ashida:2020dkc}. Other interesting structures will also arise (e.g., a nontrivial fundamental group for the complex eigenvalue manifold, with loops of migration belonging to different homotopy equivalent classes: see e.g.~\cite{Guo:2022rza,Ding:2022juv}).
In the following we will consider, for simplicity, the simplest case where $D_p=2$.
	
In quantum mechanics, the hydrogen atom energy levels for different eigenfunctions with angular quantum numbers $(\ell,\,m)$ are degenerate because of a ``hidden'' SO(4) symmetry~\cite{Bander:1965rz}. These eigenfunctions remain distinct and admit a higher-dimensional \modi{irreducible} representation for the SO(4) group.
However, the present case is different: for a given set of parameters $\boldsymbol{p}$, \modi{the eigenfunction of the radial equation corresponding to $\omega_n$,} $\psi_{n}$, is completely and uniquely determined by the value of $\omega_{n}$, so at the EP $\omega_{n}=\omega_{m}$ and $\psi_{n}=\psi_{m}$. 
Because of this degeneracy between the eigenfunctions, the degeneracy of two QNMs does not admit an interpretation in terms of hidden symmetry.
A simple toy model to illustrate this point is discussed in the Supplemental Material~\cite{SM_Ref}. This behavior is typical of non-Hermitian systems where eigenfunctions do not form a complete set, as is the case of BH perturbation theory~\cite{Leaver:1986gd}.

\noindent \textbf{\em Waveform at resonance.}
The existence of EPs has important experimental applications in optics~\cite{2024ApPhL.124f0502M,2015Natur.525..354Z,PhysRevLett.103.093902,2013PhRvA..88e3810G,Ruter:2010qjb,2021Sci...372...88O}, acoustics~\cite{2018PhRvL.121h5702D}, and condensed matter physics~\cite{2021Sci...372...88O}.
These applications rely on controlling the migration of the relevant parameters. This is not possible in an astrophysical setting.
In addition, applications in quantum physics rely on the probabilistic interpretation of the wavefunction, which does not apply in the present context. 
How would a resonance appear in gravitational-wave signals?	
In classical mechanics, a driving force oscillating at exactly the same frequency as the normal modes of the system produces a new solution whose amplitude grows linearly in time, gradually destabilizing the system. We expect (and indeed, we will find) a similar behavior in our case.

Let us neglect for simplicity the (weak) dependence on the initial data in the QNM ringing and write it as $\sum_{n}E_{n}\ee^{-\ii\omega_{n} t}$, where $E_{n}$ denotes the excitation factors~\cite{Leaver:1986gd,Berti:2006wq,Zhang:2013ksa,Oshita:2021iyn,Oshita:2025ibu}:
\begin{equation}\label{excitation_factor}
  E_{n}=-\dfrac{A_{\text{out}}(\omega_{n},p_{i})}{2\ii\omega_{n}A_{\text{in}}\p(\omega_{n},p_{i})}\,.
\end{equation}
For $\boldsymbol{p}\approx \boldsymbol{p_{\star}}$ we have $A_{\text{in}}(\omega) \approx f(\omega)(\omega-\omega_{n})(\omega-\omega_{m})$, where we omit the dependence on $p_{i}$ for simplicity. Thus, when $\delta\omega_{nm}=\omega_{n}-\omega_{m}$ is small enough, we have $A_{\text{in}}\p(\omega_{n})=f(\omega_{n})\delta\omega_{nm}\approx-A_{\text{in}}\p(\omega_{m})$, and therefore $E_{n}\approx -E_{m}$, which is consistent with Ref.~\cite{Motohashi:2024fwt}. When we add the two contributions to the waveform, since $\omega_{n}\simeq \omega_{m}$, we get
\begin{equation}~\label{eq:two_modes_contribution}
  \begin{aligned}
    E_{n}\ee^{-\ii\omega_{n}t}+E_{m}\ee^{-\ii(\omega_{n}-\delta\omega_{nm})t}&\approx\modi{-2i\sin\left(\dfrac{\delta\omega_{nm}t}{2}\right)E_{n}\ee^{-\ii\omega_{n}t}}\\
     &\approx \modi{\left[\dfrac{A_{\text{out}}(\omega_{n})}{2\omega_{n}f(\omega_{n})}\right]\times t\ee^{-\ii\omega_{n}t}\,,}
  \end{aligned}
\end{equation}
where we used Eq.~(\ref{excitation_factor}) and the expansion of $A_{in}(\omega)$. This approximation is valid as long as $\delta\omega_{nm}/\text{Im}(\omega_{n})\ll1$.
Therefore a resonant signal has a characteristic linear growth in time due to the beating between the two different frequencies.
Unlike the harmonic oscillator analog in classical mechanics, the linear growth does not cause any instability, as the exponential damping eventually dominates over the linear growth. 
	
The more careful treatment in the Supplemental Material~\cite{SM_Ref} shows that for arbitrary initial data and small enough $\delta\omega_{nm}$, the resonant waveform has the form
\begin{equation}\label{resonance_signal}
  \begin{aligned}
    \psi_{s}=&A_r\ee^{\ii m\phi}\left\{ (u-u_0)\ee^{-\ii \omega_n u}  {}_s S_{\ell m}(a \omega_n, \theta) \right.\\
             &\quad\quad\quad\left.+\ii\ee^{-\ii \omega_n u}\dfrac{\dd {}_s S_{\ell m}}{\dd \omega}(a \omega_n, \theta)\right\}\,.\\
  \end{aligned}
\end{equation}
Here \modi{$u=t-x$,} $x$ is the constant radial position of the observer,
$u_0$ and $A_{r}$ are constants depending on the initial data and the position of the observer, and all the derivatives are evaluated at $\omega=\omega_{n}$. 
The resonance between two nearly coincident modes produces two terms that are not captured by ``ordinary'' waveforms: (1) the term found by the simple qualitative analysis presented above, which has the ordinary angular dependence but amplitude growing linearly in time $t$, and (2) an exponentially decaying term with a different angular dependence. Note that the term proportional to $u_{0}$ is nothing but the ordinary QNM ringing term due to $\omega_{n}$, while the part proportional to $u$ is characteristic of the resonance. 

\noindent \textbf{\em Fitting the resonance.}
We now show that the omission of the resonant term in the fit of a resonant waveform would induce a large bias in the inferred QNM frequency. 

We compare fits performed with two different waveform models. The first is just a superposition of damped sinusoids,
\begin{equation}\label{oldform}
  h_{\text{DS}}(t)=\text{Re}\left(\sum_{i=1}^{n} A_{i}\ee^{-\ii\omega_{i}t}\right) \,,
\end{equation}
while the second contains also a resonant term:
\begin{equation}\label{resform}
  h_{\text{R}}(t)=\text{Re}\left(A_{\text{R}}t\ee^{-\ii\omega_{1}t}+\sum^{n}_{i=1} A_{i}\ee^{-\ii\omega_{i}t}\right) \,.
\end{equation}
We define a mismatch function in terms of the inner product $(f,g)=\int_{t_{\text{start}}}^{t_{\text{end}}} f(t)g(t)\dd t$ as follows:
\begin{equation}\label{mismatch}
  \mathcal{M}=\dfrac{(h-\psi,h-\psi)}{(\psi,\psi)}\,.
\end{equation}

As a concrete example, consider the resonance due to adding a ``bump'' perturbation to the ordinary Regge-Wheeler potential $V_{\text{RW}}$~\cite{Regge:1957td} that characterizes axial (i.e., odd parity) perturbations of the Schwarzschild spacetime~\cite{Cheung:2021bol,Motohashi:2024fwt,Yang:2024vor}:
\begin{equation}\label{bump}
  V(x)=V_{\text{RW}}(x)+\epsilon\ee^{-(x-d)^2} \,,
\end{equation}
where $x=r+2\ln(r-2)$ is the tortoise coordinate and the ``bump'' has two free parameters: the amplitude $\epsilon$ and the distance $d$ between the bump and the peak of the RW potential. \modi{Such a ``bump'' can be generated, e.g., by a thin mass shell~\cite{Laeuger:2025zgb}.} 

\begin{figure}
  \centering
  \includegraphics[width=1.0\linewidth]{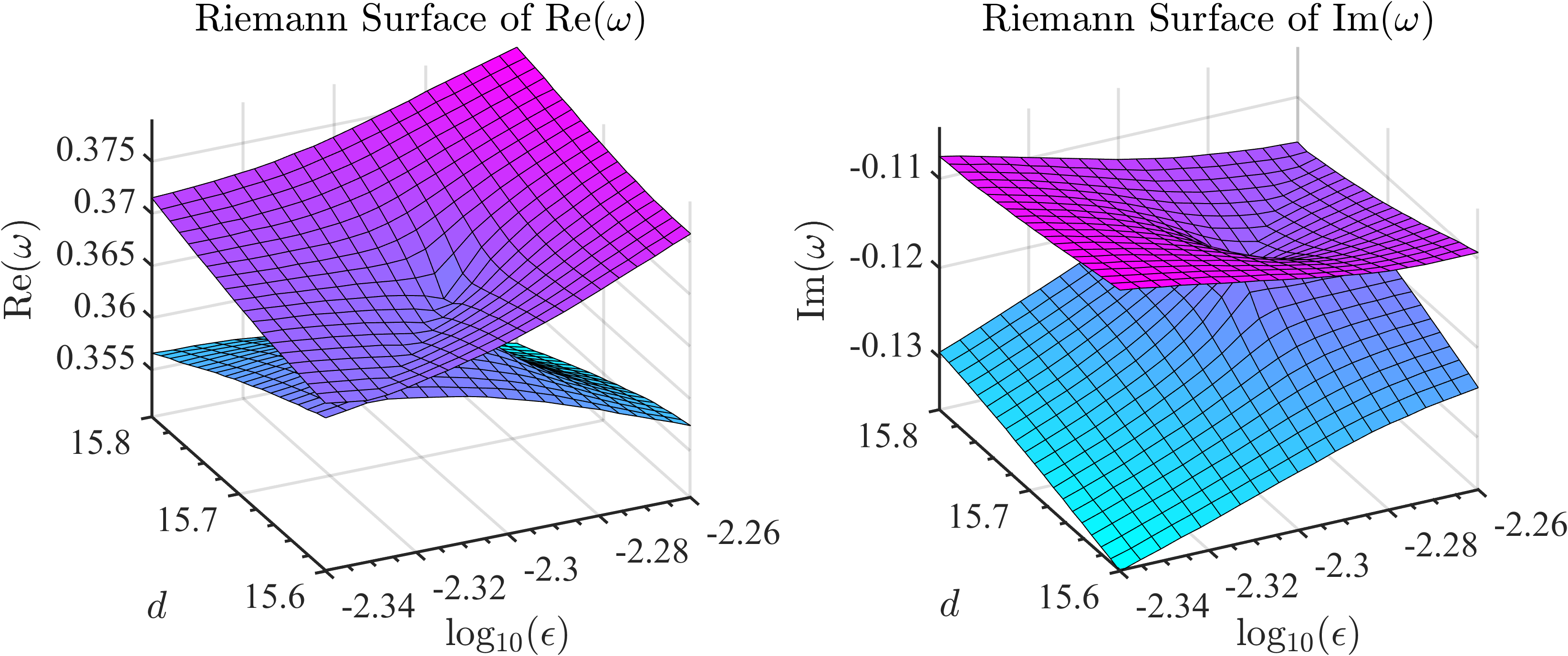}
  \caption{Riemann surface structure of the QNMs around the EP for the bump potential \modi{in Eq.~(\ref{bump})}. As in Fig.~\ref{fig:riemannsurface}, the real part and the imaginary part are displayed separately. The resonance between the fundamental mode and the first overtone occurs at $\epsilon_{\star}=10^{-2.294}\approx0.005$ and $d_{\star}=15.698$, where the modes coalesce into $\omega_{\star}\approx0.365-0.117\ii$.}
  \label{fig:bumpresonance}
\end{figure}

By integration with a shooting method, we find a resonance between the fundamental mode and the first overtone with frequency $\omega_{\star}\approx0.365-0.117\ii$ when $\epsilon=\epsilon_{\star}=10^{-2.294}\approx0.005$ and $d=d_{\star}=15.698$. The Riemann surface around this EP is shown in Fig.~\ref{fig:bumpresonance}.
    
\begin{figure*}
  \centering
  \includegraphics[width=1.0\linewidth]{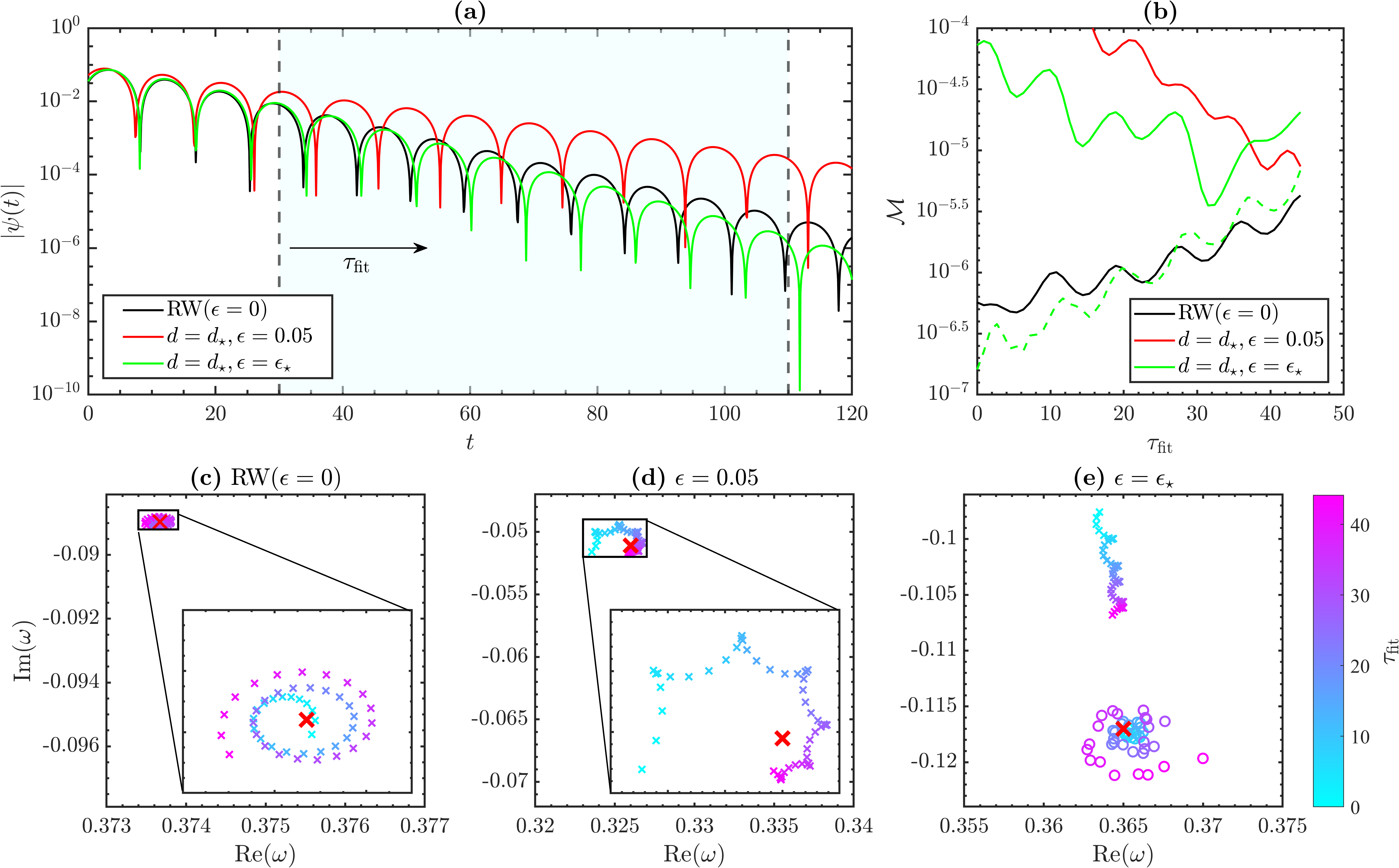}
  \caption{Time-domain waveforms and fitting results in three different cases. (a) Upper left: ringdown signal for the unperturbed RW potential (black), a perturbative bump with $\epsilon=0.05\approx10\epsilon_{\star}$ and $d=d_{\star}$ (red), and for bump parameters fine tuned to resonance (green). We have chosen the time origin $t=0$ (somewhat arbitrarily) to exclude the initial transient part and focus on the ringdown. 
  The vertical dashed lines are the boundaries of the fitting region, which ranges from $t=30$ ($\tau_\mathrm{fit}=0$) to $t=110$. 
    (b) Upper right: mismatch as a function of the starting time of the fit, $\tau_{\text{fit}}=t_{\mathrm{start}}-30$. Solid lines refer to the waveforms in panel~(a), fitted using a single QNM waveform; the green dashed line refers to a fit of the resonant waveform with a resonant ringdown signal.
    Bottom panels: frequencies inferred by fitting the corresponding signals. The red cross marks the fundamental mode. Colored crosses/circles are the frequencies found by varying the starting time of the fit from $t_{\mathrm{start}}=30$ or $\tau_{\text{fit}}=0$
    (light blue) to
     around $t_{\mathrm{start}}=75$ or $\tau_{\text{fit}}=45$
     (purple) for the RW potential in panel~(c), the non-resonant bump in panel~(d), and the resonant bump in panel~(e).
    In panel~(e), the crosses are frequencies inferred by fitting a single QNM to the resonant signal, while the circles were found by fitting a resonant waveform model.
    }
  \label{fig:signal}
\end{figure*}

We find the time-domain signal by numerically solving the perturbation equation. We compare three different signals: one for the RW potential (where no EP exists), one for the RW potential with a bump ``close'' to the EP but not quite coincident with it ($\epsilon=0.05\approx10\epsilon_{\star}$ and $d=d_{\star}$), and one corresponding exactly to the EP of the ``RW plus bump'' potential.

In panel~(a) of Fig.~\ref{fig:signal} we show the results. In all three cases the observer is located at $x=200$, and the initial data consists of a Gaussian wavepacket
localized around $x\approx10$.
The \modi{full} ringdown signal of a BH with a perturbing bump consists of two parts: the prompt ringdown and the ``echoes'' (waves reflected back and forth between the bump and the peak of the RW potential).
The \modi{full} ringdown signal
\modi{can be} described by a superposition of the destabilized QNMs only after the echoes are observed, while the prompt ringdown is nothing but the original signal produced by the unperturbed RW potential, with only small deviations due to the transmission by the bump~\cite{Barausse:2014tra,Berti:2022xfj,Yang:2024vor}. Since we are interested in the effect of resonances on the ringdown signal, we only need to analyze the part of the signal that is indeed characterized by the destabilized spectrum. This is the blue
\modi{``echoes region''}
delimited by vertical dashed lines in panel~(a). 
We perform the fitting within this region by allowing for different starting times. 
We vary the starting time from $\tau_{\text{fit}}=0$ (i.e. $t_{\mathrm{start}}=30\approx2d_{\star}$, corresponding roughly to the time at which the reflected wave carrying the desired information about the destabilized spectrum first reaches the observer~\cite{Yang:2024vor}) to $\tau_\mathrm{fit}=45$. We stop the fit somewhat arbitrarily at $t_\mathrm{end}=110$, when the signal amplitude is very small (typically of order $10^{-6}$).
    
In panel~(b) of Fig.~\ref{fig:signal} we show the mismatch as a function of the starting time of the fit. The fit of the resonant waveform with a resonant model (dashed green) performs much better than the fit with a single exponential (solid green), with mismatches comparable to the mismatch achieved by fitting exponentials to the unperturbed RW case. 
    
In the bottom panels we extract the fundamental mode frequency by fitting in the three different cases. A large red cross marks the theoretical value of the frequency; smaller crosses (circles) represent fits with the pure damped sinusoids of Eq.~\eqref{oldform} or with the resonant waveform of Eq.~\eqref{resform}, respectively, with colors indicating the starting time of the fit.
In the unperturbed RW and non-resonant bump cases of panels (c) and (d), the fitted frequencies deviate by less than $0.05\%$ (for RW) and $1\%$ (non-resonant bump) from the expected value. In the presence of the bump, the first overtone is more slowly damped, and it has a larger effect on the early-time fit.
In panel~(e) we observe that by fitting a resonant signal with a simple damped sinusoid the deviations from the ``true'' frequency are as large as $17\%$ at early fitting times, and they never quite converge even at late times.
The outcome improves dramatically if we instead fit the signal with a resonant waveform: the fitted frequency has deviations below $5\%$ at early times, and it rapidly converges to the ``correct'' value.

In the Supplemental Material~\cite{SM_Ref} we repeat the fits using multiple overtones, further confirming that the model with a linear growth in time provides a better fit of resonant waveforms.

\modi{We have demonstrated the effect of the resonance in the time domain by a specific toy model, but the behavior predicted in Eq.~(\ref{eq:two_modes_contribution}) is completely general as long as the QNMs are solutions of Eq.~(\ref{qnmcondition}). If the resonance affects modes with long enough damping, an analysis based on a pure superposition of damped exponentials will be inadequate. For example, in theories beyond general relativity one generally finds a coupled system of perturbation equations, and resonances could amplify the observability of non-tensorial degrees of freedom. This was recently demonstrated in the specific case of Einstein-Maxwell-axion theory~\cite{Takahashi:2025uwo}.}

\noindent \textbf{\em Conclusions.}
In this Letter we considered resonances in which two QNMs are either completely degenerate (EPs) or exhibit avoided crossings. We demonstrated that near EPs, the two modes are just different covers of the same complex function on a Riemann surface. We also demonstrated that the characteristic time-domain resonant signal has an amplitude that grows linearly in time, just like harmonic oscillators driven by a resonant force. However, in the BH case, the resonance originates from beating frequencies, and it does not cause an instability due to the dominance of exponential damping at late times.
We also demonstrated numerically that models that include the linear growth in time are much better at fitting resonant waveforms. 
The relevance of these considerations in the context of modified gravity theories and gravitational-wave phenomenology is an interesting topic for future work.

\noindent \textbf{\em Acknowledgments.}
Y.Y and N.F.~would like to thank the Johns Hopkins
University for hospitality during the early stages of this work. Y. Y. would like to thank Lijing Shao and the members of his group for the fruitful discussions and unwavering support during the later stages of this work. 
N.F. acknowledges funding from the FCT Grant Agreement No.~2023.06263.CEECIND/CP2830/CT0004 and support to the Center for Astrophysics and
 Gravitation (CENTRA/IST/ULisboa) through FCT Grant No.~UID/99/2025. 
E.B. is supported by NSF Grants No.~AST-2307146, No.~PHY-2513337, No.~PHY-090003, and No.~PHY-20043, by NASA Grant No.~21 ATP21-0010, by John Templeton Foundation Grant No.~62840, by the Simons Foundation [MPS-SIP-00001698, E.B.], by the Simons Foundation International [SFI-MPS-BH-00012593-02], and by Italian
 Ministry of Foreign Affairs and International Cooperation
 Grant No.~PGR01167.

\noindent \text{\em Data availability.} The data are available from the
 authors upon reasonable request.

\section*{Supplemental Material}
\noindent \textbf{\em Proof of the properties around the exceptional point.}
For a system with $D_p$ free parameters $ p_{i} $ ($i=1,\dots, D_p$), the modified Teukolsky equation can be cast in the form
\begin{equation}\label{radialequation}
	\left[\partial_{x}^{2}+\omega^{2}-V(p_{i},\omega, x)\right]\Psi=0 \,.
\end{equation}
\modi{Here and below we omit the indices $(\ell,\,m)$ to simplify the notation.} We define two linearly independent Jost solutions, $ \psi_{\omega} $ and $ \phi_{\omega} $, such that	
\begin{equation}\label{psi}
	\psi_\omega(x) \rightarrow \left\{
	\begin{aligned}
		&\ee^{-i\omega x}, &&x\rightarrow-\infty \,, \\
		& A_{\text{in}}(\omega) \ee^{-i \omega x}+A_{\text{out}}(\omega) \ee^{i \omega x}, &&x \rightarrow +\infty  \,,
	\end{aligned}
	\right.
\end{equation}

\begin{equation}\label{phi}
	\phi_\omega(x) \rightarrow \left\{
	\begin{aligned}
		& B_{\text{in}}(\omega) \ee^{i \omega x}+B_{\text{out}}(\omega)\ee^{-i \omega x}, &&x \rightarrow -\infty  \,, \\
		&\ee^{i\omega x}, &&x\rightarrow+\infty \,,
	\end{aligned}
	\right.
\end{equation}
where we dropped the dependence on the parameters $p_i$ for simplicity. By definition, QNMs are just the first-order poles of the analytic function $ A_{\text{in}}(\omega, p_{i}) $:
\begin{equation}\label{QNMcondition}
	A_{\text{in}}(\omega_{n},p_{i})=0\,.
\end{equation}

In general, given a QNM $\omega_{n}$, for infinitesimal changes in the parameters one has
\begin{equation}
	\dd A_{\text{in}}=\nabla_{\boldsymbol{p}}A_{\text{in}}\cdot\dd \boldsymbol{p}+\dfrac{\partial A_{\text{in}}}{\partial \omega}\dd \omega_{n}=0\,.
\end{equation}
With $\omega_{n}=\omega_{n}(p_{i})$ we find
\begin{equation}\label{migration_SM}
	\nabla_{\boldsymbol{p}}\omega_{n}
	=-\dfrac{1}{A_{\text{in}}^{\prime}(\omega_{n},p_{i})}\nabla_{\boldsymbol{p}}A_{\text{in}}(\omega_{n},p_{i})\,,
\end{equation}
where a prime denotes a derivative with respect to $\omega$.
At the EP, the two modes $ \omega_{n} $ and $ \omega_{m} $ coincide: $\omega_n = \omega_m \equiv \omega_\star$. For the region around the EP, the two modes are so close that they nearly form a second-order zero, or explicitly
\begin{equation}\label{nearexpansion}
	A_{\text{in}}(p_{i},\omega)=f(\omega,p_{i})\left[\omega-\omega_{n}(p_{i})\right]\left[\omega-\omega_{m}(p_{i})\right] \,,
\end{equation}
where $ f(\omega_{n})\approx f(\omega_{m})\approx f(\omega_{\star}) $. Let us define $ \delta\omega_{nm}=\omega_{n}-\omega_{m} $. By replacing Eq.~(\ref{nearexpansion}) into Eq.~(\ref{migration_SM}) for both $ \omega_{n} $ and $\omega_{m}$, we get

\begin{equation}\label{quadraticmigration}
	\delta\omega_{nm}\dd \delta\omega_{nm}=\dfrac{1}{f(\omega_{\star})}\dd \boldsymbol{p}\cdot\left[ \nabla_{\boldsymbol{p}}A_{\text{in}}(\omega_{n},p_{i})+\nabla_{\boldsymbol{p}}A_{\text{in}}(\omega_{m},p_{i})\right]\,.
\end{equation}
Notice that $\delta\omega_{nm}$ is a function of all the $ p_{i} $'s. For a small region around the resonance point, we thus obtain
\begin{equation}
	\delta\omega_{nm}=2\sqrt{\dfrac{1}{f(\omega_{\star})} (\boldsymbol{p}-\boldsymbol{p_{\star}})\cdot\nabla_{ \boldsymbol{p}}A_{\text{in}}(\omega_{\star},p_{i\star})}\,.
	\label{eq:riemann}
\end{equation}

This derivation only requires $ A_{\text{in}}(\omega, p_{i}) $ to be an analytic function of more than two free parameters, so it is quite general.

\noindent \textbf{\em A toy model for the exceptional point.}
The general properties of EPs in non-Hermitian systems can be illustrated with a very simple model. Consider the Hamiltonian~\cite{Ding:2022juv,Heiss:2012dx,Ashida:2020dkc}
\begin{equation}\label{toy_matrix}
	H(z)=\begin{pmatrix}
		0 & 1 \\
		1 & z
	\end{pmatrix} \,,
\end{equation}
that can represent a coupled two-level system.
\modi{For $z\in \mathbb{C}$ the Hamiltonian will be non-Hermitian, and the system will depend on two real parameters $p_1=\text{Re}(z)$ and $p_2=\text{Im}(z)$, using the notation of the main text and of Fig.~1 there.} The two eigenvalues are
\begin{equation}\label{toy_eigen}
	\omega_{\pm}(z)=\dfrac{1}{2}(z\pm\sqrt{4+z^2})\,.
\end{equation}

The square root of a complex function in Eq.~(\ref{toy_eigen}) is the one-to-two map plotted in Fig.~1 of the main text. 
The complex plane is split into two Riemann sheets connected by some choice of branch cut. The endpoint of the branch cut is then an EP in non-Hermitian physics. 
In this toy model, the two EPs are located at $z=\pm2\ii$, where the eigenvalues $\omega_{\pm}$ are identical. From Fig.~1 of the main text we also see that when $z$ varies continuously along a loop around one of the EPs, the eigenvalues will circle around the Riemann surface twice and get back to their original value.

If we slightly displace $z$ from one of the EPs, so that $z\approx\pm2\ii(1+\dfrac{1}{2}\delta z)$, we find $\delta\omega_{\pm}=\omega_{+}-\omega_{-}\approx\pm2\sqrt{\delta z}$.
Therefore the presence of two Riemann sheets is a generic property of any EP formed by two coalescing eigenvalues. We can also compute the eigenvectors (up to a constant) of this matrix: 
\begin{equation}
	\psi_{\pm}=\begin{pmatrix}
		1 \\
		\omega_{\pm}(z)
	\end{pmatrix}
\end{equation}
The eigenvectors will become degenerate at the EP. Since different eigenvectors are orthogonal to each other, their norm must satisfy $|\psi_{\pm}|^{2}\rightarrow 0$ when $z\rightarrow\pm 2\ii$.
Thus, the situation is quite different from those degeneracies that imply some hidden symmetry of the system, as in the case of degeneracies between different bands in a Hermitian system, where two distinct bands intersect and the corresponding energy eigenfunctions merge into a larger-dimensional irreducible representation of another symmetry group at a point with higher symmetry. The EP instead leads to a degeneracy between the eigenfunctions, and it does not admit the same interpretation. 

\begin{figure}[t]
	\centering
	\includegraphics[width=0.9\linewidth]{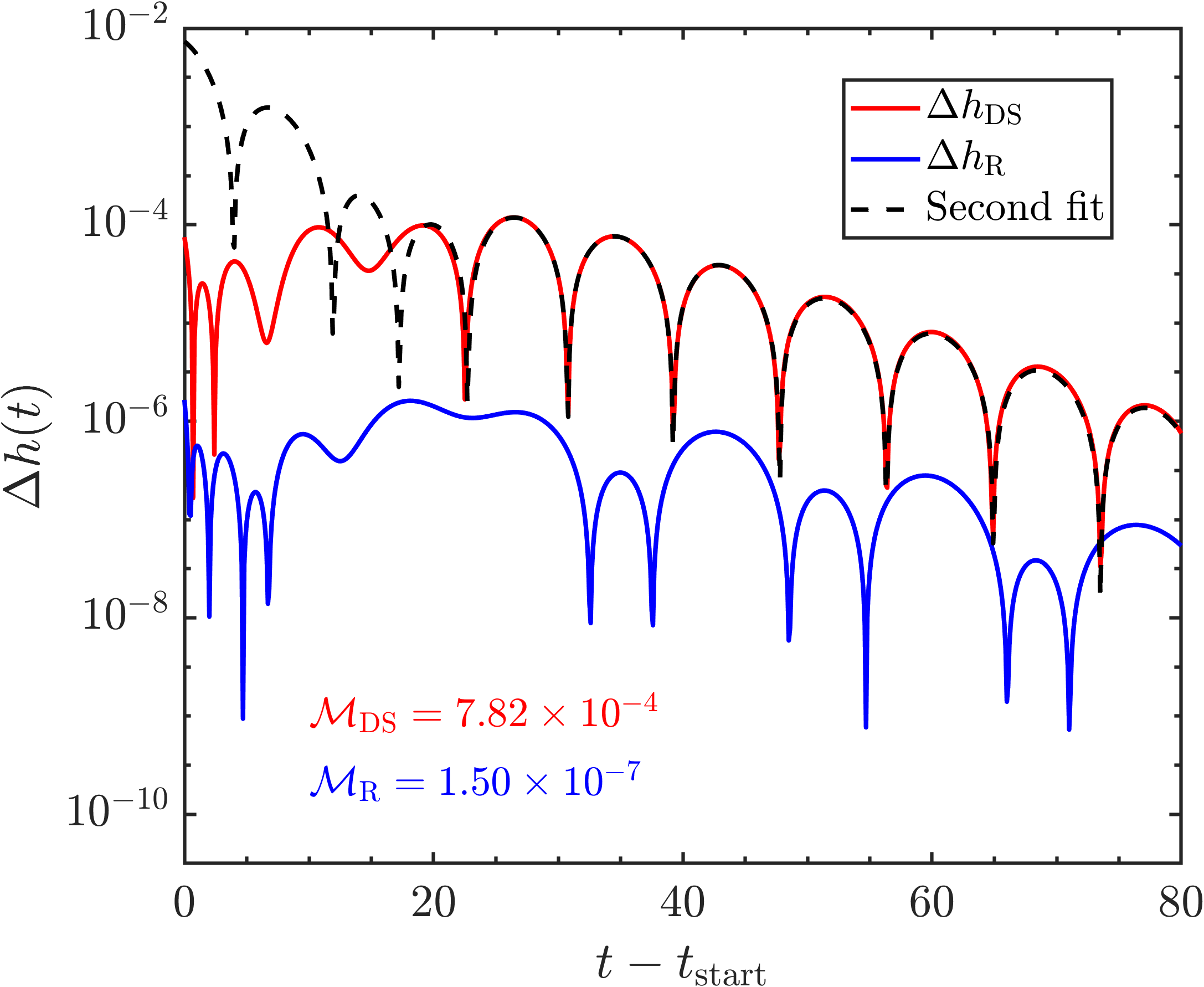}
	\caption{Results of fitting a resonant waveform $\psi$ using resonant and non-resonant waveforms with $n=4$. The red line is the residual $\Delta h_{\text{DS}}=\psi-h_{\text{DS}}(t)$ between the resonant waveform and a model with only damped sinusoids, while the blue line is the corresponding residual $\Delta h_{\text{R}}=\psi-h_{\text{R}}(t)$ for the resonant waveform model. The black dashed line is the result of fitting $\Delta h_{\text{DS}}$ again with $h_{\text{R}}(t)$.}
	\label{fig:restfit}
\end{figure}

\noindent \textbf{\em A multi-mode fit of resonant waveforms.}
As further proof that a model with linear growth in time is indeed a good representation of the time domain signal, here we extend the analysis in the main text by fitting the signal with multiple overtones (up to $n=4$). We fix all frequencies to their theoretical value and we treat the two degenerate modes as a single mode with frequency $\omega_{1}$, so $\omega_{2}$ is actually the second overtone.
In Fig.~\ref{fig:restfit} we plot the difference between the resonant signal $\psi(t)$ and the fit $h_{i}$, $\Delta h_{i}(t)=\psi(t)-h_{i}(t)$, where $h_i=h_\textrm{DS}$ refers to simple damped sinusoids, while $h_i=h_\textrm{R}$ refers to the resonant model.
The mismatch using damped sinusoids is $\mathcal{M}_{\text{DS}}=7.82\times 10^{-4}$, much higher than the mismatch of $\mathcal{M}_{\text{R}}=1.50\times 10^{-7}$ found by using the resonant waveform. We also perform a second fit of the ``residual'' $\Delta h_{\text{DS}}(t)$ using the resonant waveform with $n=1$. It is clear from the black dashed line in Fig.~\ref{fig:restfit} that this second fit matches $\Delta h_{\text{DS}}$ very well at late times, but not at early times. This is because fitting with higher overtones will suppress the mismatch to some extent but only at early times, as higher overtones rapidly damp to very small magnitudes.

\noindent \textbf{\em Proof of the time dependence of resonant waveforms.}
The Green's function for the Teukolsky equation can be written as
\begin{equation}
	\begin{aligned}
		G_{s}&(t,\boldsymbol{x}|t\p,\boldsymbol{x}\p)=\\
		&\int_{\mathscr{C}}\dfrac{\dd p}{2\pi \ii}\ee^{p(t-t\p)}\sum_{\ell m}\widetilde{G}^{\ell m}_{s}(x,x\p){}_{s}S_{\ell m}(p,\theta)_{s}S_{\ell m}(p,\theta\p )\ee^{\ii m(\phi+\phi\p )}\,,
	\end{aligned}
\end{equation}
where $p=-i\omega$ is the Laplace parameter and $\widetilde{G}^{\ell m}_{s}(x,x\p)$ is the radial Green's function. For initial data $ \psi_{s}(t=0,\boldsymbol{x})=f(\boldsymbol{x}) $ and $ \dot{\psi_{s}}(t=0,\boldsymbol{x})=g(\boldsymbol{x}) $, the solution reads
\begin{equation}
	\psi_{s}=\sum_{\ell m}\int_{\mathrm{C}}\dfrac{\dd p}{2\pi \ii }\ee^{pt}\left[\int\dd x\p \widetilde{G}^{\ell m}_{s}(x,x\p )\mathscr{J}^{\ell m}_{s}(x\p )\right]{}_{s}S_{\ell m}(p,\theta)\ee^{\ii m\phi}\,,
\end{equation}                                                          
where
\begin{equation}
	\mathscr{J}^{\ell m}_{s}(x\p)=-\int \text{sin}\theta\p\dd\theta\p \dd \phi\p  \left[pf(\boldsymbol{x\p })+g(\boldsymbol{x\p })\right]{}_{s}S_{\ell m}(p,\theta\p )\ee^{\ii m\phi\p}\,.
\end{equation}

We now construct the function $ \widetilde{G}^{\ell m}_{s}(x,x\p)  $ following Ref.~\cite{Teukolsky:1973ha}. For a general BH with spin $a$ and mass $M$, the perturbation function for a spin-$s$ field in the frequency domain satisfies
\begin{equation}\label{radial}
	\Delta^{-s} \frac{\dd}{\dd r}\left(\Delta^{s+1} \frac{\dd R}{\dd r}\right)+\left(\frac{K^2-2 i s(r-M) K}{\Delta}+4 i s \omega r-\lambda\right) R=0\,,
\end{equation}
where $K\equiv(r^2+a^2)\omega-am$, $\lambda\equiv E +a^{2}\omega^{2}-2am\omega-s(s+1)$, and $E$ is the  angular eigenvalue. This equation can be rewritten as a wave equation similar to Eq.~(\ref{radialequation}):
\begin{equation}
	\begin{aligned}
		&\qquad\qquad \qquad Y_{, r * r *}+ V(r)Y=0\,,\\
		&V(r)=\dfrac{K^2-2 i s(r-M) K+\Delta(4 i r \omega s-\lambda)}{\left(r^2+a^2\right)^2}-G^2-G_{, r_{\star}}\,,
	\end{aligned}
\end{equation}
where $G=s(r-M)/(r^{2}+a^{2})+r\Delta/(r^2+a^2)^2$~\cite{Teukolsky:1973ha}. In terms of the Jost solutions of Eqs.~(\ref{psi}) and (\ref{phi}), the Green's function for $ Y(r) $ reads
\begin{equation}
	\widetilde{G}_{Y}(x,x\p )=-\dfrac{\phi(x)\psi(x^{\prime})\mathscr{\theta}(x-x\p )+\psi(x)\phi(x^{\prime})\mathscr{\theta}(x\p -x)}{2i\omega A_{in}(\omega)}\,.
\end{equation}
By direct algebra (and omitting some indices for brevity) we get the Green's function of $ \psi_{s} $ for $x>x\p\gg1$:
\begin{equation}
	\begin{aligned}
		\widetilde{G}^{\ell m}_{s}(x,x\p )&=-\dfrac{1}{2i\omega A_{\text{in}}(\omega)}\dfrac{(\Delta\p)^{s+1}}{r^{\prime 2}+a^{2}}\left[\Phi(x)\Psi(x\p )\right]\\
		&\approx-\dfrac{1}{2i\omega}\dfrac{\ee^{i\omega x}}{r^{2s+1}}\left[ \dfrac{\ee^{-i\omega x\p}}{(r\p)^{1-s}}+\dfrac{A_{\text{out}}}{A_{\text{in}}}\dfrac{\ee^{i\omega x\p}}{(r\p)^{s+1}}\right]\,,
	\end{aligned}
\end{equation}
where $\Phi$ is a solution of Eq.~(\ref{radial}) that is purely outgoing at spatial infinity, while $\Psi$ is purely ingoing at the horizon.

Now define the quantity
\begin{equation}\label{amplitude}
	\widetilde{K}^{n}=-\dfrac{A_{\text{out}}(\omega_{n})}{2i\omega_{n} A_{\text{in}}\p(\omega_{n})}\int\dd x\p\dfrac{\ee^{i\omega_{n} x\p}}{(r\p)^{s+1}} \mathscr{J}^{\ell m}_{s}(\omega_{n},x\p)\,,
\end{equation}
which is what we usually regard as the amplitude of the QNM. Then the full solution reads
\begin{equation}\label{qnmfunction}
	\psi_{s}=\sum_{\ell mn}\dfrac{\ee^{-i\omega_{n} (t-x)}}{r^{2s+1}}\widetilde{K}^{n}{}_{s}S_{\ell m}(a\omega_{n},\theta)\ee^{\ii m\phi}\,.
\end{equation}

Close to the EP $\omega_{n}=\omega_{m}=\omega_{\star}$, we can hardly distinguish the two modes. When we add them together we have
\begin{equation}
	\begin{aligned}
		\psi_{s}=\dfrac{\ee^{\ii m\phi}}{r^{2s+1}}&\left[\widetilde{K}^{n}{}_{s}S_{\ell m}(a\omega_{n},\theta)\ee^{-i\omega_{n} (t-x)}\right.\\
		\quad\quad&\left.+\widetilde{K}^{m}{}_{s}S_{\ell m}(a\omega_{m},\theta)\ee^{-i\omega_{m} (t-x)}\right]\,.
	\end{aligned}
\end{equation}

If $\delta\omega_{nm}$ is very small, we can Taylor expand $A_{\text{in}}(\omega)\approx f(\omega)(\omega-\omega_{n})(\omega-\omega_{m})$ for $\omega\simeq \omega_{\star}$, so that $A_{\text{in}}\p(\omega_{n})=f(\omega_{n})\delta\omega_{nm}$ and $A_{\text{in}}\p(\omega_{m})=-f(\omega_{m})\delta\omega_{nm}$. Thus we may expect that $\widetilde{K}^{n}$ and $-\widetilde{K}^{m}$ will be close to each other. In general they will not be, especially when the initial data are very far away from the horizon, as can be seen from Eq.~(\ref{amplitude}). However we can always decompose $\widetilde{K}^{m}=-\widetilde{K}^{n}+\Delta\widetilde{K}$, so that
\begin{equation}\label{signal}
	\begin{aligned}
		\psi_{s}=\dfrac{\ee^{\ii m\phi}}{r^{2s+1}}&\left\{\widetilde{K}^{n}\delta\omega_{nm}\left[ \dfrac{\dd {}_s S_{\ell m}}{\dd \omega}- \ii (t - x)  {}_s S_{\ell m}(a \omega_n, \theta)\right]\ee^{-i \omega_n (t - x)} \right.\\
		&\quad\quad\quad\left.+\Delta \widetilde{K}{}_s S_{\ell m}(a \omega_n, \theta)\ee^{-i \omega_n (t - x)}\right\}\,.\\
	\end{aligned}
\end{equation}

The term on the second line is nothing but the QNM with frequency $\omega_{n}$ ringing with a different amplitude. The term on the first line has amplitude of order unity:
\begin{equation}\label{Iamplitude}
	\ii A_{r}\equiv \widetilde{K}^{n}\delta\omega_{nm}=-\dfrac{1}{r^{2s+1}}\dfrac{A_{\text{out}}(\omega_n)}{2i\omega_n f(\omega_n)}\int\dd x\p\dfrac{\ee^{i\omega_n x\p}}{(r\p)^{s+1}} \mathscr{J}^{\ell m}_{s}(\omega_n,x\p)\,.
\end{equation}

By defining $u=t-x$ and rearranging some terms, the waveform can be written in a more compact form:
\begin{equation}
	\begin{aligned}
		\psi_{s}=&A_r\ee^{\ii m\phi}\left\{ (u-u_0)\ee^{-\ii \omega_n u}  {}_s S_{\ell m}(a \omega_n, \theta) \right.\\
		&\quad\quad\quad\left.+\ii\ee^{-\ii \omega_n u}\dfrac{\dd {}_s S_{\ell m}}{\dd \omega}(a \omega_n, \theta)\right\}\,,\\
	\end{aligned}
\end{equation}
where the second line in Eq.~(\ref{signal}) has been absorbed into the ``starting time'' $u_0$.

\bibliography{refs}

\end{document}